\documentclass{article}

\usepackage{enumerate}
\usepackage[a4paper]{geometry}
\usepackage{amssymb,amsmath,mathtools,xcolor,graphicx,xspace,colortbl,ragged2e,rotating} %
\usepackage{amsfonts}  
\usepackage{amsmath}  
\usepackage{amssymb}  
\usepackage{amsthm}  
\usepackage{color}  
\usepackage{graphicx}  
\usepackage{mathtools}  
\usepackage{hyperref}  
\usepackage{tikz}
\usetikzlibrary{positioning, shapes.geometric}

\providecommand{\U}[1]{\protect\rule{.1in}{.1in}}
\allowdisplaybreaks
\newtheorem{theorem}{Theorem}

\newtheorem{definition}[theorem]{Definition}
\newtheorem{example}[theorem]{Example}

\newtheorem{notation}[theorem]{Notation}
\newtheorem{problem}[theorem]{Problem}
\newtheorem{proposition}[theorem]{Proposition}
\newtheorem{remark}[theorem]{Remark}

\usepackage{cite}

\begin{document}

\def\sres{\operatorname*{sres}}

\title{A Basis-preserving Algorithm for Computing \\the B\'ezout Matrix of Newton Polynomials}

\author{Jing Yang, Wei Yang\thanks{%
Corresponding author.}\\[5pt]
SMS--HCIC--School of Mathematics and Physics,\\
Center for Applied Mathematics of Guangxi,\\
Guangxi Minzu University, Nanning 530006, China\\
yangjing0930@gmail.com; weiyang020499@163.com}
\date{}
\maketitle
\begin{abstract}
This paper tackles the problem of constructing B\'ezout matrices for Newton polynomials in a basis-preserving approach that operates directly with the given Newton basis, thus avoiding the need for transformation from Newton basis to monomial basis. This approach significantly reduces the computational cost and also mitigates numerical instability caused by basis transformation. For this purpose, we investigate the internal structure of B\'ezout matrices in Newton basis and design a basis-preserving algorithm that generates the B\'ezout matrix in the specified basis used to formulate the input polynomials. Furthermore, we show an application of the proposed algorithm on constructing confederate resultant matrices for Newton polynomials. Experimental results demonstrate that the proposed methods perform superior to the basis-transformation-based ones.


\end{abstract}

\section{Introduction}
Resultant theory is a crucial component of computer algebra and has received extensive research in both theoretical and practical aspects  (just list a few
\cite{1853_Sylvester,2003_Lascoux_Pragacz,2004_Toca_Vega,1967_collins,
1971_barnett,1989_Ho,2006_Li,
2008_Terui,
2017_Bostan_Andrea_Krick,
2021_Hong_Yang,
2021_Cox_DAndrea,
2023_Hong_Yang,
2021_Hong_Yang_sub}). Most of the studies so far have focused on polynomials in standard basis. However, with the growing popularity of basis-preserving algorithms in various applications  \cite{1987_Farouki_Rajan,1991_Goodman_Said,1993_Carnicer_Pena,2006_Amiraslani_Corless,2003_Delgado_Pena}, there is a need to study resultants and subresultant polynomials for polynomials in non-standard basis  (see \cite{2004_Bini_Gemignani,2007_Marco_Martinez,2007_Aruliah_Corless_Gonzalez_Vega_Shakoori,2022_Wang_Yang}). This paper aims to address this need by presenting an approach for constructing B\'ezout matrices of Newton polynomials, which covers a wide class of polynomials in the interpolation problem.
Standard polynomials in monomial basis can also be viewed as a specialization of Newton polynomials (if we do not require that the nodes in Newton basis are distinct).  In the settings of the paper, the
input polynomials are formulated in  Newton basis and so are the output B\'ezout matrix.

This study is motivated by the observation that the basis-preserving  B\'ezout matrix usually has a simpler form, compared with the  B\'ezout matrix formulated in monomial basis. A natural way to construct  the B\'ezout matrix of two Newton polynomials is: (1) to convert the polynomials into expressions in monomial basis, (2) construct their  B\'ezout matrix in monomial basis, and (3) change the  B\'ezout matrix from monomial basis
to Newton basis. It is seen that
two basis transformations are involved, which often causes a high computational cost and results in numerical instability as well. 

To address these problems caused by basis transformation, we present a basis-preserving algorithm for computing the  B\'ezout matrix for two Newton polynomials. More explicitly, the algorithm takes two univariate polynomials in Newton basis as input,
and  produces a B\'ezout matrix for the given polynomials, while ensuring that the resulting matrix is formulated in the given basis used to define the input polynomials. To achieve this goal, we investigate the internal structure hidden in B\'ezout matrices in Newton basis, based on which a basis-preserving algorithm is designed for constructing such a matrix. Notably, the proposed algorithm does not require any basis transformation, thus avoiding the heavy computational cost and numerical instability caused by such transformations. We also demonstrate an application of the proposed algorithm in the construction of the confederate resultant matrix for Newton polynomials. Experimental results show that the proposed algorithms perform significantly better than the basis-transformation-based ones.

The current work can be viewed as a refinement of the work in \cite{2001_Yang}, where the B\'ezout matrix in general basis is considered  and various notable properties of this matrix are developed. However, from the computational view, it is almost impossible to develop a uniform and efficient algorithm to compute  B\'ezout matrices in an arbitrarily given general basis,  since it heavily relies on the structure of the chosen basis. Thus we need to develop specialized algorithms for specific instances of general basis.
Newton basis is a typical one with nice properties and wide applications. However, as far as we know, there is no work on the construction of  B\'ezout matrix for Newton polynomials. We hope the finding in this paper can help people have a better understanding on the structure of B\'ezout matrix in Newton basis and promote its application on formulating various resultant matrices and subresultant polynomials.   

The paper is structured as follows. In Section \ref{sec:preliminaries}, we first review the concepts of B\'ezout matrix in a given basis, which is followed by a formal statement of the addressed problem in Section \ref{sec:problem}.
Section \ref{sec:main_result} is devoted to present the  main result of the paper (see Theorem \ref{thm:main} and the algorithm ${\sf BezNewton\_preserving}$).
The correctness of the main theorem and the algorithm as well as the complexity of the proposed algorithm are verified in Section \ref{sec:proof}. In Section \ref{sec:comparison}, we present a detailed comparison on the proposed algorithm and the basis-transformation algorithm both in computational complexity and practical performance.
 In Section \ref{sec:application}, we show an application of the proposed algorithm in the construction of confederate resultant matrices for Newton polynomials. The paper is concluded in Section \ref{sec:conclusion} with some further remarks.

\section{Preliminaries}\label{sec:preliminaries}

In this section, we briefly review the concept of the B\'ezout matrix for two univariate polynomials, which is formulated for an arbitrarily given basis (where a non-standard basis is allowed). Throughout the paper, we assume $\mathbb{F}$ to be the fractional field of an integral domain. 

The B\'ezout matrix of two given polynomials is formulated through their Cayley quotient. Consider $F,G\in\mathbb{F}[x]$ with degrees $n$ and $m$, respectively, where $n\ge m$. The Cayley quotient of $F$ and $G$ is defined as
\begin{align}\label{delta}
\Delta (x,y):=\frac{\Lambda (x,y)}{x-y}
\end{align}
where
\begin{align}\label{lamda}
\Lambda(x,y)=\begin{vmatrix}
        F(x)  & F(y) \\
        G(x) & G(y)
\end{vmatrix}
\end{align} 
It is noted that $\Lambda(x,y)=0$ when $x=y$. Thus the polynomial $\Lambda(x,y)$ has a factor $x-y$, which implies that $\Delta(x,y)$  is essentially a  polynomial in $\mathbb{F}[x]$. One may also note the fact that
\[
\deg(\Delta,x)=\deg(\Delta,y)=n-1
\] 

Let $\mathbb{F}_{n-1}[x]$ be the vector space consisting of all the polynomials with degree no greater than $n$ and $\boldsymbol{\Phi}_{n-1}(x)=(\phi_{n-1}(x),\ldots,\phi_{0}(x))^T$ be a basis of $\mathbb{F}_{n-1}[x]$. Then we can rewrite $\Delta(x,y)$ into an expression formulated in terms of $\phi_{ij}(x,y)$'s where $\phi_{ij}(x,y)=\phi_i(x)\phi_j(y)$, that is
\begin{align*}
\Delta(x,y)=\sum_{i=1}^{n}\sum_{j=1}^{n}b_{ij}\phi_{n-i}(x)\phi_{n-j}(y)
\end{align*}
With the help of expressions for $\Delta(x,y)$, we construct the B\'ezout resultant matrix of $F$ and $G$ in the basis $\boldsymbol{\Phi}_{n-1}$.
\begin{definition}[B\'ezout matrix \cite{2007_Aruliah_Corless_Gonzalez_Vega_Shakoori}]
Let $\boldsymbol{\Phi}_{n-1}(x)=(\phi_{n-1}(x),\ldots,\phi_{0}(x))^T$ be a basis of $\mathbb{F}_{n-1}[x]$. The B\'ezout resultant matrix of $F$ and $G$ with respect to $x$ in the basis $\boldsymbol{\Phi}_{n-1}$ is defined as an $n\times n$ matrix $\boldsymbol{B}_{\boldsymbol{\Phi}}(F,G)$ such that
\begin{align*}
\Delta(x,y)=\boldsymbol{\Phi}_{n-1}(x)^{T}\boldsymbol{B}_{\boldsymbol{\Phi}}(F,G)\boldsymbol{\Phi}_{n-1}(y)
\end{align*}
\end{definition}

\begin{remark}We make the following remarks:
\begin{enumerate}[(1)]
\item When $F$ and $G$ are clear from the context, we can abbreviate $\boldsymbol{B}_{\boldsymbol{\Phi}}(F,G)$ as $\boldsymbol{B}_{\boldsymbol{\Phi}}$.
\item Let $\boldsymbol{\Psi}(x)=(\psi_{n-1}(x),\ldots,\psi_{0}(x))$ be another basis of $\mathbb{F}_{n-1}[x]$. Then there exists a transition matrix $\boldsymbol{U}$ such that $\boldsymbol{\Phi}=\boldsymbol{U}\boldsymbol{\Psi}$. Note that $\Delta$ is independent of the chosen basis. Thus it can be further derived that $\boldsymbol{B}_{\boldsymbol{\Psi}}=\boldsymbol{U}^T\boldsymbol{B}_{\boldsymbol{\Phi}}\boldsymbol{U}$. In other words, the B\'ezout matrix of two polynomials in different bases are congruent.

\item When $\boldsymbol{\Phi}(x)=(x^{n-1},\ldots,x^0)$, the B\'ezout matrix of  $F$ and $G$ is exactly the well known B\'ezout matrix with respect to $x$.

\item It is known that the classical B\'ezout matrix, i.e., the B\'ezout matrix in monomial basis, is symmetric, and it can be further deduced from (1) and (2) that the B\'ezout matrix in any basis is symmetric.

\item In the rest of the paper, we always assume that $F$ and $G$ are polynomials in terms of $x$. For the sake of simplicity, we will not specify that their B\'ezout matrix is with respect to the variable $x$.
\end{enumerate}
\end{remark}

The main goal of this paper is to present a basis-preserving method for constructing the B\'ezout matrix of two Newton polynomials where the resulting matrix is required to be formulated in the given basis. To give a formal statement of the problem addressed in the paper, we need to recall the concepts of Newton basis and Newton polynomials, which are the basic concepts in many textbooks on numerical analysis (e.g., \cite{2002_Stoer}).

\begin{definition}\label{def:Newton_basis}
Given $\boldsymbol{\lambda}=(\lambda_1,\ldots,\lambda_s)\in\mathbb{F}^{s}$, let $\boldsymbol{N}_{\boldsymbol{\lambda}}(x)=(N_{s}(x),\ldots,N_{1}(x),N_{0}(x))^{T}$, where
$$
N_{i}(x)=\left\{
\begin{array}{ll}
1 & \text{for~}i=0; \\
(x-\lambda_{i})N_{i-1}(x)  & \text{for~}i>0.
\end{array}
\right.
$$
We call $\boldsymbol{N}_{\boldsymbol{\lambda}}(x)$ the Newton basis of $\mathbb{F}_s[x]$ with respect to $\boldsymbol{\lambda}$ and a linear combination of $N_i(x)$'s is called a Newton polynomial.
\end{definition}
It is obvious that any polynomial in $\mathbb{F}_{s}[x]$ can be written into an equivalent Newton polynomial. Hence, for $F,G\in\mathbb{F}[x]$ with degrees $n$ and $m $ respectively, where $n\ge m$, 
we assume that $F$ and $G$ are given in their Newton form where the Newton basis is $\boldsymbol{N}_{\boldsymbol{\lambda}}(x)=(N_{n}(x),\ldots,N_{1}(x),N_{0}(x))^{T}$ with $\boldsymbol{\lambda}\in\mathbb{F}^n$. More explicitly, we assume 
\begin{equation}\label{eq:F+G}
F\left ( x \right ) =\sum_{i=0}^{n} a_{i} N_{i}(x),\quad G\left ( x \right ) =\sum_{i=0}^{m} b_{i} N_{i }(x).
\end{equation}
It should be pointed out that the Newton basis used to formulate the B\'ezout matrix of $F$ and $G$  is a subset of $\boldsymbol{N}_{\boldsymbol{\lambda}}$ obtaining by truncating the first polynomial, rather than the basis itself. In the remaining part of the paper, we denote the truncated $\boldsymbol{N}_{\boldsymbol{\lambda}}$ 
with $\tilde{\boldsymbol{N}}_{\boldsymbol{\lambda}}$ for simplicity.

Next we clarify why we are interested in the B\'ezout matrix of $F$ and $G$ in the Newton basis  $\tilde{\boldsymbol{N}}_{\boldsymbol{\lambda}}$ through an illustrative example below.

\begin{example}\label{ex:comparison_Bez}
Given the Newton basis with respect to $\boldsymbol{\lambda}=(-1,0,2)$, that is, 
$$\boldsymbol{N}_{\boldsymbol{\lambda}}(x)=(N_{3}(x),N_{2}(x),N_{1}(x),N_{0}(x))^T$$
where
\[
N_{0}(x)=1, \quad N_{1}(x)=x+1,\quad N_{2}(x)=x(x+1),
\quad N_{3}(x)=x(x+1)(x-2),
\]
let
$F$ and $G$ be polynomials in  $\boldsymbol{N}_{\boldsymbol{\lambda}}$. More explicitly,
\begin{align}
F&=a_3N_{3}(x)+a_2N_{2}(x)+a_1N_{1}(x)+a_0N_{0}(x),\label{eq:F}\\
G&=b_2N_{2}(x)+b_1N_{1}(x)+b_0N _{0}(x).\label{eq:G}
\end{align}
First, we calculate that
\begin{align*}
\Delta(x,y)=&\ \ \ \ a_{3}b_{2}N_{2}(x)N_{2}(y)+a_{3}b_{1}N_{1}(x)N_{2}(y)+a_{3}b_{0}N_{0}(x)N_{2}(y)+a_{3}b_{1}N_{2}(x)N_{1}(y)\\
&+(a_{3}b_{0}-2a_{3}b_{1}+a_{2}b_{1}-a_{1}b_{2})N_{1}(x)N_{1}(y)+(-3a_{3}b_{0}+a_{2}b_{0}-a_{0}b_{2})N_{0}(x)N_{1}(y)\\
&+a_{3}b_{0}N_{2}(x)N_{0}(y)+(-3a_{3}b_{0}+a_{2}b_{0}-a_{0}b_{2})N_{1}(x)N_{0}(y)\\
&+(-a_{2}b_{0}+a_{0}b_{2}+3a_{3}b_{0}+a_{1}b_{0}-a_{0}b_{1})N_{0}(x)N_{0}(y).
\end{align*}
Thus the B\'ezout matrix  of $F$ and $G$  in  $\tilde{\boldsymbol{N}}_{\boldsymbol{\lambda}}$  is
\begin{equation}\label{eq:bez_nb}
\boldsymbol{B}_{\tilde{\boldsymbol{N}}_{\boldsymbol{\lambda}}}=\begin{pmatrix}
        a_{3}b_{2} & a_{3}b_{1} & a_{3}b_{0} \\
        a_{3}b_{1} & a_{3}b_{0}-2a_{3}b_{1}+a_{2}b_{1}-a_{1}b_{2} & -3a_{3}b_{0}+a_{2}b_{0}-a_{0}b_{2} \\
        a_{3}b_{0} & -3a_{3}b_{0}+a_{2}b_{0}-a_{0}b_{2} & -a_{2}b_{0}+a_{0}b_{2}+3a_{3}b_{0}+a_{1}b_{0}-a_{0}b_{1}
\end{pmatrix}.
\end{equation}
We may further expand $\Delta(x,y)$ into an expression in the monomial basis and obtain
\begin{align*}
\Delta(x,y)=\,&\ \ \ a_{3}b_{2}x^{2}y^{2}+(a_{3}b_{2}+a_{3}b_{1})xy^{2}+(a_{3}b_{1}+a_{3}b_{0})y^{2}\\
&+(a_{3}b_{2}+a_{3}b_{1})x^{2}y+(a_{3}b_{2}+a_{3}b_{0}+a_{2}b_{1}-a_{1}b_{2})xy+Q_{1}y\\
&+(a_{3}b_{1}+a_{3}b_{0})x^{2}+Q_{1}x+Q_{2},
\end{align*}
where\begin{align*}
&Q_{1}=\,-a_{3}b_{1}-a_{3}b_{0}+a_{2}b_{1}-a_{1}b_{2}+a_{2}b_{0}-a_{0}b_{2},\\
&Q_{2}=\,-2a_{3}b_{1}-2a_{3}b_{0}+a_{2}b_{1}-a_{1}b_{2}+a_{2}b_{0}-a_{0}b_{2}+a_{1}b_{0}-a_{0}b_{1}.\end{align*}
Then the B\'ezout matrix  of $F$ and $G$  in  monomial basis  is
\begin{equation}\label{eq:bez_pb}
\boldsymbol{B}_{\tilde{\boldsymbol{P}}}=\begin{pmatrix}
        a_{3}b_{2} & a_{3}b_{2}+a_{3}b_{1} & a_{3}b_{1}+a_{3}b_{0}  \\
        a_{3}b_{2}+a_{3}b_{1} & a_{3}b_{2}+a_{3}b_{0}+a_{2}b_{1}-a_{1}b_{2} & Q_{1} \\
        a_{3}b_{1}+a_{3}b_{0} & Q_{1} & Q_{2} 
\end{pmatrix}.
\end{equation}
\end{example}

By comparing \eqref{eq:bez_nb} and \eqref{eq:bez_pb} in Example \ref{ex:comparison_Bez}, we have the following observations:
\begin{itemize}
\item The B\'ezout matrix of Newton polynomials in monomial basis is more complicated than that in the given Newton basis. It can be verified through the two following matrices which consist of the numbers of terms in the entries of  $\boldsymbol{B}_{\tilde{\boldsymbol{N}}_{\boldsymbol{\lambda}}}$ 
and $\boldsymbol{B}_{\tilde{\boldsymbol{P}}}$:
\begin{align*}
\#\ \text{terms in the entries of }\boldsymbol{B}_{\tilde{\boldsymbol{N}}_{\boldsymbol{\lambda}}}&\ =\ \begin{pmatrix}
1&1&1\\
1&4&3\\
1&3&5
\end{pmatrix},\\
\#\ \text{terms in the entries of }\boldsymbol{B}_{\tilde{\boldsymbol{P}}}&\ =\ \begin{pmatrix}
1&2&2\\
2&4&6\\
2&6&8
\end{pmatrix}
\end{align*}

\item In fact, the basis-preserving formulation of B\'ezout matrix can preserve the structure in the B\'ezout matrix better. For example, the $(2,2)$-entry in $\boldsymbol{B}_{\tilde{\boldsymbol{N}}_{\boldsymbol{\lambda}}}$ can be obtained by subtracting the $(1,2)$-entry multiplied by $2$ from the $(1,3)$-entry and adding $a_{2}b_{1}-a_{1}b_{2}$.
Such patterns are buried when the B\'ezout matrix is formulated in monomial basis.
\end{itemize}

\section{Problem Statement}\label{sec:problem}

Now we are ready to give a formal statement of the problem we address in this paper.

\begin{problem}\label{problem}
The formulation problem of B\'ezout matrix in Newton basis is stated as:
\begin{description}
\item[In\ \ :] $\boldsymbol{\lambda}\in\mathbb{F}^n$, which determines a Newton basis $\boldsymbol{N}_{\boldsymbol{\lambda}}(x)=(N_{n}(x),\ldots,N_0(x))^T$ of $\mathbb{F}_{n}[x]$;

$F,G\in\mathbb{F}[x]$ with degrees $n$ and $m$ respectively, where $n\ge m$, and written in the form:
\[F\left ( x \right ) =\sum_{i=0}^{n} a_{i} N_{i}(x),\quad G\left ( x \right ) =\sum_{i=0}^{m} b_{i} N_{i }(x).\]
\item[Out:] $\boldsymbol{B}_{\tilde{\boldsymbol{N}}_{\boldsymbol{\lambda}}}$, the B\'ezout matrix of $F$ and $G$ in the basis $\tilde{\boldsymbol{N}}_{\boldsymbol{\lambda}}=(N_{n-1},\ldots,N_0)^T$.
\end{description}
\end{problem}

One straightforward way to solve Problem \ref{problem} is: (1) converting the input polynomials into expressions in monomial basis, (2) constructing the B\'ezout matrix in monomial basis, and (3) changing back to the given basis. However, this naive approach involves two basis transformations which take expensive costs. It can be shown that the computational complexity of the above method is $\mathcal{O}(n^3)$. Readers may refer to Subsection \ref{sub:complexity} for its detailed complexity analysis. On the other hand,   
Chionh et al. proposed a recursive way for constructing the B\'ezout resultant in monomial basis with a complexity of $\mathcal{O}(n^2)$. One may naturally wonder whether it is possible to formulate the B\'ezout matrix of Newton polynomials in the given Newton basis with the complexity $\mathcal{O}(n^2)$.

Another disadvantage of the above straightforward approach is the back-and-forth basis transformation often cause numerical instability, especially when the values of $\lambda_i$'s are big. It motivates us to investigate a basis-preserving algorithm for formulating the B\'ezout matrix in Newton basis, which does not involve any basis transformation and thus can avoid the numerical instability issue.

\section{ Main Results}\label{sec:main_result}
\noindent We introduce the following short-hand notation for a concise presentation of the main theorem.
\begin{notation}
$[i,j]=\begin{vmatrix}
                a_{i} & a_{j}\\
                b_{i} & b_{j}
        \end{vmatrix}.$
\end{notation}
\begin{theorem}\label{thm:main}
Given $\boldsymbol{\lambda}\in\mathbb{F}^n$ which determines a Newton basis $\boldsymbol{N}_{\boldsymbol{\lambda}}(x)$ of $\mathbb{F}_{n}[x]$ and
$F,G\in\mathbb{F}_{n}[x]$ be as in \eqref{eq:F+G}, let $\boldsymbol{B}_{\tilde{\boldsymbol{N}}_{\boldsymbol{\lambda}}}=(c_{i,j})_{n\times n}$ be the B\'ezout matrix of F and G in the Newton basis $\tilde{\boldsymbol{N}}_{\boldsymbol{\lambda}}$. Then the entries in $\boldsymbol{B}_{\tilde{\boldsymbol{N}}_{\boldsymbol{\lambda}}}$ has the following relationship:
\[
c_{i,j}=c_{i-1,j+1}+(\lambda_{n-j+1}-\lambda_{n-i+2})c_{i-1,j}+[n-i+1,n-j],
\]
where
\begin{itemize}
        \item $c_{p,q}:=0$, if $p = 0$ or  $q = n+1$;
        \item  $b_{p}:=0$, if $p>m$;
        \item $\lambda_p:=0$, if $p>n$.  
\end{itemize}
\end{theorem}
\begin{example}[Continued from Example \ref{ex:comparison_Bez}]
Recall that $\boldsymbol{\lambda}=(-1,0,2)$. By Theorem \ref{thm:main},
we have
\begin{itemize}
\item when $(i,j)=(1,2)$,
\[
c_{1,2}=c_{0,3}+(\lambda_{2}-\lambda_{4})c_{0,2}+[3,1]=a_{3}b_{1}-a_{1}b_{3}=a_{3}b_{1},
\]
since $c_{0,3}=c_{0,2}=b_3=0$;

\item when $(i,j)=(2,2)$,
\begin{align*}
c_{2,2}=\,&c_{1,3}+(\lambda_{2}-\lambda_{3})c_{1,2}+[2,1]\\
=\,&a_3b_0+(0-2)a_3b_1+a_2b_1-a_1b_2\\
=\,&a_3b_0-2a_3b_1+a_2b_1-a_1b_2.
\end{align*}
\end{itemize} 
Both agree with the expressions of $c_{1,2}$ and $c_{2,2}$ in Example \ref{ex:comparison_Bez}.

\end{example}

Now we present the following algorithm for constructing the B\'ezout matrix of two Newton polynomials with basis unchanged.

\bigskip
\noindent\textbf{Algorithm.} $\boldsymbol{B}_{\tilde{\boldsymbol{N}}_{\boldsymbol{\lambda}}}\leftarrow {\sf BezNewton\_preserving}(F,G)$
\begin{enumerate}
\item[In\ \ :]  $\boldsymbol{\lambda}\in\mathbb{F}^n$, which determines a Newton basis $\boldsymbol{N}_{\boldsymbol{\lambda}}(x)$ of $\mathbb{F}_{n}[x]$;\\ 
$F,G\in\mathbb{F}[x]$ in Newton basis $\boldsymbol{N}_{\boldsymbol{\lambda}}(x)$ with degrees $n$ and $m$ respectively, where $n\ge m$, i.e.,
\begin{eqnarray*}
F\left ( x \right ) =\sum_{i=0}^{n} a_{i} N_{i}(x),\quad G\left ( x \right ) =\sum_{i=0}^{m} b_{i} N_{i }(x) .
\end{eqnarray*}
\item[Out:] the B\'ezout matrix $\boldsymbol{B}_{\tilde{\boldsymbol{N}}_{\boldsymbol{\lambda}}}$ of $F$ and $G$ in $\tilde{\boldsymbol{N}}_{\boldsymbol{\lambda}}$.
\end{enumerate}
\begin{enumerate}[Step $1$.]
\item  Initialization.

$(c_{i,j})_{n\times (n+1)}\leftarrow \begin{pmatrix}
        [n,n-1] & \cdots  & [n,0] & 0\\
        & \ddots  & \vdots & \vdots\\
        &  & [1,0] & 0
\end{pmatrix}$
\item Recursion.\\
For $i=2\ldots,n$ \\
\hspace*{0.5cm} For $j=i,\ldots,n$ \\
\hspace*{1.2cm}$c_{i,j}\gets c_{i,j}+c_{i-1,j+1}+(\lambda_{n-j+1}-\lambda_{n-i+2})c_{i-1,j}$

\item Symmetrization.\\
For $i=1\ldots,n$ \\
\hspace*{0.5cm} For $j=i+1,\ldots,n$ \\
\hspace*{1.2cm}$c_{j,i}\gets c_{i,j}$

\item Truncation.\\
$
\boldsymbol{B}_{\tilde{\boldsymbol{N}}_{\boldsymbol{\lambda}}}\leftarrow\begin{pmatrix}
c_{1,1} & \cdots  & c_{1,n}\\
\vdots  & \ddots & \vdots   \\
c_{n,1} & \cdots &c_{n,n}
\end{pmatrix}_{n\times n}
$
\item Return $\boldsymbol{B}_{\tilde{\boldsymbol{N}}_{\boldsymbol{\lambda}}}$.
\end{enumerate}

\begin{example}[Continued from Example \ref{ex:comparison_Bez}]
Let $F$ and $G$ be as in \eqref{eq:F} and \eqref{eq:G}.
Now we show how to construct the B\'ezout matrix  of $F$ and $G$  in  $\tilde{\boldsymbol{N}}_{\boldsymbol{\lambda}}$ by using the algorithm ${\sf BezNewton\_preserving}(F,G)$. Following the steps described in the algorithm, we have
{\footnotesize
\begin{align*}
&\xrightarrow{\text{Initialization}}
\begin{pmatrix}
        [3,2] & [3,1] & [3,0] &0\\
        &  [2,1]& [2,0]&0\\
        &  & [1,0] &0
\end{pmatrix} \\
&\xrightarrow{\text{Recursion:}\ i=2}
\begin{pmatrix}
        [3,2] & [3,1] & [3,0] &0\\
         & [2,1]+[3,0]+(\lambda_{2}-\lambda_{3})[3,1]& [2,0]+0+(\lambda_{1}-\lambda_{3})[3,0]&0\\
         &  & [1,0] &0
\end{pmatrix}\\ 
&\xrightarrow{\text{Recursion:}\ i=3}\setlength{\arraycolsep}{4pt} 
\begin{pmatrix}
        [3,2] & [3,1] & [3,0] &0\\
         &  [2,1]+[3,0]+(\lambda_{2}-\lambda_{3})[3,1]& [2,0]+0+(\lambda_{1}-\lambda_{3})[3,0]&0\\
         &  & [1,0]+0+(\lambda_{1}-\lambda_{2})([2,0]+0+(\lambda_{1}-\lambda_{3})[3,0])&0
\end{pmatrix}\\ 
&\xrightarrow{\text{Symmetrization}}\setlength{\arraycolsep}{4pt} \begin{pmatrix}
        [3,2] & [3,1] & [3,0] &0\\
        [3,1] &  [2,1]+[3,0]+(\lambda_{2}-\lambda_{3})[3,1]& [2,0]+(\lambda_{1}-\lambda_{3})[3,0]&0\\
        [3,0] & [2,0]+(\lambda_{1}-\lambda_{3})[3,0] & [1,0]+(\lambda_{1}-\lambda_{2})[2,0]+(\lambda_{1}-\lambda_{2})(\lambda_{1}-\lambda_{3})[3,0]&0
\end{pmatrix}.\\
&\xrightarrow{\text{Truncation}}
\setlength{\arraycolsep}{4pt} 
\boldsymbol{B}_{\tilde{\boldsymbol{N}}_{\boldsymbol{\lambda}}}\leftarrow\begin{pmatrix}
        [3,2] & [3,1] & [3,0] \\
        [3,1] &  [2,1]+[3,0]+(\lambda_{2}-\lambda_{3})[3,1]& [2,0]+(\lambda_{1}-\lambda_{3})[3,0]\\
        [3,0] & [2,0]+(\lambda_{1}-\lambda_{3})[3,0] & [1,0]+(\lambda_{1}-\lambda_{2})[2,0]+(\lambda_{1}-\lambda_{2})(\lambda_{1}-\lambda_{3})[3,0]
\end{pmatrix}.
\end{align*}}
The output of the algorithm is $\boldsymbol{B}_{\tilde{\boldsymbol{N}}_{\boldsymbol{\lambda}}}$.

It should be mentioned that in order to show what is exactly going on in the execution of the algorithm, we keep $\boldsymbol{\lambda}$ in the parametric form. In fact, we can evaluate $\boldsymbol{B}_{\tilde{\boldsymbol{N}}_{\boldsymbol{\lambda}}}$ at  $\boldsymbol{\lambda}=(-1,0,2)$, which gives us the same expression as in \eqref{eq:bez_nb}.  
\end{example}

\begin{proposition}\label{prop:algorithm}
\
\begin{enumerate}[(1)]
\item The algorithm ${\sf BezNewton\_preserving}(F,G)$ terminates at a finite number of steps, and the algorithm output is correct.

\item The complexity of the algorithm  ${\sf BezNewton\_preserving}$ is $\mathcal{O}(n^2)$.
\end{enumerate}
\end{proposition}

\section{Proofs}\label{sec:proof}
\subsection{Proof of Theorem \ref{thm:main}}
\begin{proof}
By the multi-linearity of determinants, we simplify the expression of $\Lambda \left ( x,y \right )$ in \eqref{lamda} and derive the following: 
\begin{align}\label{lamda=abb}
\Lambda \left ( x,y \right ) =\det
\begin{bmatrix}
                \sum\limits_{i=0 }^{n} a_{i} N_{i}(x)  & \sum\limits_{i=0}^{n} a_{i} N_{i}(y)\\
                \sum\limits_{i=0}^{m} b_{i} N_{i}(x) &\sum\limits_{i=0}^{m} b_{i} N_{i}(y)
        \end{bmatrix}
         =\sum_{0\le i \le  n \atop 0\le j \le n} \det
         \begin{bmatrix}
         a_{i} & a_{j}\\
         b_{i} & b_{j}
 \end{bmatrix}N_{i}(x)N_{j}(y)
\end{align}
(where $b_{i}=0$ if $i>m$.)

On the other hand,  $\Lambda(x,y)$ can be obtained from the multiplication of $\Delta\left ( x,y \right )$ and $x-y$. More specifically,
we have\begin{align} \label{eqs:lambdabb}
\Lambda \left ( x,y \right ) =\,&\Delta \left ( x,y \right ) (x-y) \nonumber\\
=\,&\tilde{\boldsymbol{N}}_{\lambda}(x)^{T}\boldsymbol{B}_{\tilde{\boldsymbol{N}}_{\boldsymbol{\lambda}}}\tilde{\boldsymbol{N}}_{\lambda}(y)(x-y) \nonumber\\ \nonumber
=\,&\begin{pmatrix}
        N_{n-1}(x) &
        \cdots &
        N_{1}(x)&
        N_{0}(x)\end{pmatrix} \begin{pmatrix}
        c_{1,1} & \cdots  & c_{1,n}\\
        \vdots  & \ddots  & \vdots\\
        c_{n,1} & \cdots &c_{n,n}
\end{pmatrix}\begin{pmatrix}
        N_{n-1}(y) \\
        \vdots \\
        N_{1}(y)\\
        N_{0}(y)\end{pmatrix}(x-y)\nonumber\\ 
=\,&(x-y)\sum\limits_{1\le i,j\le n}c_{i,j}N_{n-i}(x)N_{n-j}(y)
\end{align}
Next we rewrite \eqref{eqs:lambdabb} into an expression in terms of $N_i(x)N_j(y)$'s where $0\le i,j\le n$. 

Note that
\begin{align}
&x\sum\limits_{1\le i,j\le n}c_{i,j}N_{n-i}(x)N_{n-j}(y)\notag\\
=&\sum\limits_{1\le i,j\le n}c_{i,j}(x-\lambda_{n-i+1}+\lambda_{n-i+1})N_{n-i}(x)N_{n-j}(y)\notag\\
=&\sum\limits_{1\le i,j\le n}\left(c_{i,j}(x-\lambda_{n-i+1})N_{n-i}(x)N_{n-j}(y)+c_{i,j}\lambda_{n-i+1}N_{n-i}(x)N_{n-j}(y)\right)\notag\\
=&\sum\limits_{1\le i,j\le n}c_{i,j}N_{n-i+1}(x)N_{n-j}(y)+\sum\limits_{1\le i,j\le n}c_{i,j}\lambda_{n-i+1}N_{n-i}(x)N_{n-j}(y)\label{eqs:x*Lambda}
\end{align}
With similar deduction, we obtain
\begin{align}
&y\sum\limits_{1\le i,j\le n}c_{i,j}N_{n-i}(x)N_{n-j}(y)\notag\\
=&\sum\limits_{1\le i,j\le n}c_{i,j}N_{n-i}(x)N_{n-j+1}(y)+\sum\limits_{1\le i,j\le n}c_{i,j}\lambda_{n-j+1}N_{n-i}(x)N_{n-j}(y)\label{eqs:y*Lambda}
\end{align}
The substraction of \eqref{eqs:y*Lambda} from \eqref{eqs:x*Lambda} yields
\begin{align}
\Lambda(x,y)=\,&x\sum\limits_{1\le i,j\le n}c_{i,j}N_{n-i}(x)N_{n-j}(y)-y\sum\limits_{1\le i,j\le n}c_{i,j}N_{n-i}(x)N_{n-j}(y)\notag\\ \nonumber
=\,&\sum\limits_{1\le i,j\le n}c_{i,j}N_{n-i+1}(x)N_{n-j}(y)+\sum\limits_{1\le i,j\le n}c_{i,j}\lambda_{n-i+1}N_{n-i}(x)N_{n-j}(y)\\ \nonumber
&-\left(\sum\limits_{1\le i,j\le n}c_{i,j}N_{n-i}(x)N_{n-j+1}(y)+\sum\limits_{1\le i,j\le n}c_{i,j}\lambda_{n-j+1}N_{n-i}(x)N_{n-j}(y)\right)\\ \nonumber
=&\sum\limits_{0\le i,j\le n}c_{i+1,j}N_{n-i}(x)N_{n-j}(y)+\sum\limits_{0\le i,j\le n}c_{i,j}\lambda_{n-i+1}N_{n-i}(x)N_{n-j}(y)\\ \nonumber
&-\left(\sum\limits_{0\le i,j\le n}c_{i,j+1}N_{n-i}(x)N_{n-j}(y)+\sum\limits_{0\le i,j\le n}c_{i,j}\lambda_{n-j+1}N_{n-i}(x)N_{n-j}(y)\right)\\ 
=&\sum\limits_{0\le i,j\le n}\left(c_{i+1,j}-c_{i,j+1}+(\lambda_{n-i+1}-\lambda_{n-j+1})c_{i,j}\right)N_{n-i}(x)N_{n-j}(y)
\label{lambda=abb}
\end{align}
(where $c_{i,j}=0$, if $i\le 0$ or $j \le 0 $.)

Comparing the coefficients of \eqref{lamda=abb} and \eqref{lambda=abb} in the term $N_{n-i}(x)N_{n-j}(y)$, we obtain
\[\begin{vmatrix}
        a_{n-i} & a_{n-j}\\
        b_{n-i} & b_{n-j}
\end{vmatrix}=c_{i+1,j}-c_{i,j+1}+(\lambda_{n-i+1}-\lambda_{n-j+1})c_{i,j}\]
which is equivalent to the following:
\begin{align*}
c_{i+1,j}=c_{i,j+1}-(\lambda_{n-i+1}-\lambda_{n-j+1})c_{i,j}+\begin{vmatrix}
                a_{n-i} & a_{n-j}\\
                b_{n-i} & b_{n-j}
        \end{vmatrix}
\end{align*}
Reset $i+1$ to be $i$ and we achieve that
\begin{align*}
c_{i,j}=c_{i-1,j+1}+(\lambda_{n-j+1}-\lambda_{n-i+2})c_{i-1,j}+[n-i+1,n-j]
\end{align*}
\end{proof}
\subsection{Proof of Proposition \ref{prop:algorithm}}

In this subsection, we show the  the algorithm  ${\sf BezNewton\_preserving}(F,G)$ terminates within finite steps and outputs the correct B\'ezout matrix. Furthermore, a detailed analysis on its computational complexity is provided.

\begin{proof}
\begin{enumerate}[(1)]
\item \emph{Termination.} The termination of the algorithm is guaranteed by the finiteness of $n$. 
\item \emph{Correctness.} The proof is given in an inductive way.
We start with the resulting matrix of the Initialization step, denoted by $M_1$. Its $(j,k)$-th entry $c_{j,k}^{(1)}$ is assigned to be $[n-j+1,n-k]$ except for those in the last column, which indicates that
the $(1,k)$-th entry in the first row is $[n,n-k]$.
When $j=1$, by Theorem \ref{thm:main},
we have
the $(1,k)$-th entry $c_{1,k}$ of $\boldsymbol{B}_{\tilde{\boldsymbol{N}}_{\boldsymbol{\lambda}}}$ where $1\le k\le n$ is: $$c_{1,k}=c_{0,k+1}+(\lambda_{n-k+1}-\lambda_{n+1})c_{0,k}+[{n,n-k}]=[n,n-k]$$
where the simplification is due to the settings $c_{0,k}=c_{0,k+1}=0$. Thus the first row of $M_1$ is the same as that of $\boldsymbol{B}_{\tilde{\boldsymbol{N}}_{\boldsymbol{\lambda}}}$.

Next we show that after the $n-1$ recursions, the $(j,k)$-th entry $c_{j,k}^{(n)}$ of the resulting matrix, denoted by $M_{n}$, is the same as that of $\boldsymbol{B}_{\tilde{\boldsymbol{N}}_{\boldsymbol{\lambda}}}$, where  $j\le n$ and $k\ge j$. 
The proof is given in an inductive way.
\begin{itemize}
\item Initial step. When $i=1$, the first row of $M_1$ is the same as that of  $\boldsymbol{B}_{\tilde{\boldsymbol{N}}_{\boldsymbol{\lambda}}}$, which is already shown above.

\item Inductive step. Assume after the $(i-1)$-th recursion, the $(j,k)$-th entry $c_{j,k}^{(i)}$ of $M_i$ is the same as the $(j,k)$-th entry $c_{j,k}^{}$  of $\boldsymbol{B}_{\tilde{\boldsymbol{N}}_{\boldsymbol{\lambda}}}$ for $j\le i$ and $k\ge j$. Now we execute the $i$-th recursion. Note that in the  $i$-th recursion, only the $(i+1)$-th row is updated. Hence
 the $(j,k)$-th entry $c_{j,k}^{(i+1)}$ of $M_{i+1}$ for $k=j+1,\ldots,n$ is
\[
c_{j,k}^{(i+1)}=\left\{
\begin{array}{ll}
c_{j,k}^{(i)}+c_{j-1,k+1}^{(i)}+(\lambda_{n-k+1}-\lambda_{n-j+2})c_{j-1,k}^{(i)}&j=i+1\\
c_{j,k}^{(1)}&j>i+1
\end{array}\right.
\]
Let $j=i+1$. Since $c_{i+1,k}^{(i)}=c_{i+1,k}^{(1)}=[n-i,n-k]$, $c_{i,k+1}^{(i)}=c_{i,k+1}$, $c_{i,k}^{(i)}=c_{i,k}$,
\[c_{i+1,k}^{(i+1)}=c_{i,k+1}+(\lambda_{n-k+1}-\lambda_{n-i+1})c_{i,k}+[n-i,n-k]\]
By Theorem \ref{thm:main},
\[c_{i+1,k}=c_{i,k+1}+(\lambda_{n-k+1}-\lambda_{n-i+1})c_{i,k}+[n-i,n-k]\]
Hence $c_{i+1,k}^{(i+1)}=c_{i+1,k}$. Note that
the first $i$-th rows of $M_{i}$  do not change in the $i$-th recursion. Therefore, the $(j,k)$-th entry $c_{j,k}^{(i+1)}$ of $M_{i+1}$ is the same as the $(j,k)$-th entry $c_{j,k}^{}$  of $\boldsymbol{B}_{\tilde{\boldsymbol{N}}_{\boldsymbol{\lambda}}}$ for $j\le i+1$ and $k\ge j$.
\end{itemize}
After $n-1$ recursions,  we obtain the upper triangular part of $\boldsymbol{B}_{\tilde{\boldsymbol{N}}_{\boldsymbol{\lambda}}}$. By the symmetry of B\'ezout matrix, we can immediately have the lower triangular part of $\boldsymbol{B}_{\tilde{\boldsymbol{N}}_{\boldsymbol{\lambda}}}$.

\item \emph{Complexity.}
The computation cost of the algorithm  ${\sf BezNewton\_preserving}$ occurs in the steps of Initialization and Recursion.
In the Initialization step, we need to compute $\frac{n(n+1)}{2}$ determinants of order $2$. Each determinant requires two multiplications and one addition. Hence the total cost in this step is $n(n+1)$ multiplications and $\frac{n(n+1)}{2}$ additions. In the Recursion step, we need to compute $\frac{n(n-1)}{2}$ entries of $\boldsymbol{B}_{\tilde{\boldsymbol{N}}_{\boldsymbol{\lambda}}}$ and each entry costs one multiplication and three additions. Hence the total cost in this step is $\frac{n(n-1)}{2}$ multiplications and $\frac{3n(n-1)}{2}$ additions.

In summary, the overall cost for one execution of the algorithm ${\sf BezNewton\_preserving}$ is $\frac{3n^2+n}{2}$ multiplications and $2n^2-n$ additions.
\end{enumerate}
\end{proof}
\section{Comparison}\label{sec:comparison}

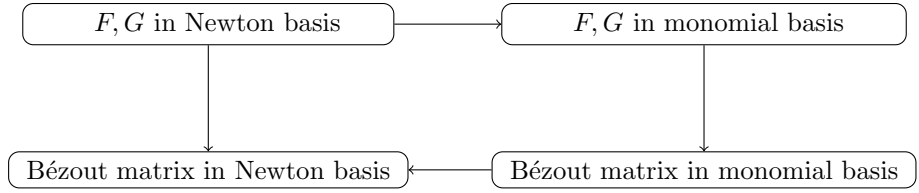
\begin{figure}
  \centering
\begin{tikzpicture}[node distance=40pt]
        \node[draw, rounded corners](start)   {\ \ \ \ \ \ \ $F,G$ in Newton basis\ \ \  \ \ \ };
        \node[draw,  rounded corners,right=of start](step 1)  {\ \ \ \ \ \ \  $F,G$  in monomial basis\ \ \ \ \ \ \  };
        \node[draw,  rounded corners,below=of start]                        (step 2)  {\ B\'ezout matrix in Newton basis \ };
        \node[draw,  rounded corners, aspect=2, below=of step 1]     (choice)  { \  B\'ezout matrix in monomial basis \ };
        \draw[->] (start)  -- (step 1);
        \draw[->] (start) -- (step 2);
        \draw[->] (step 1) -- (choice);
        \draw[->] (choice) -- (step 2);
\end{tikzpicture}
  \caption{Two approaches for computing the B\'ezout matrix in Newton basis}\label{fig:diagram}
\end{figure}

In this section, we compare the algorithm  ${\sf BezNewton\_preserving}$ for computing the B\'ezout matrix in Newton basis proposed in this paper with a basis-transformation-based algorithm, called  ${\sf BezNewton\_trans}$. The two approaches are illustrated in Figure \ref{fig:diagram}. The comparison will be carried out from two aspects: computational complexity and practical performance.  

\subsection{A description of the algorithm  ${\sf BezNewton\_trans}$ } 
For the comparison purpose, we give a brief description of the algorithm  ${\sf BezNewton\_trans}$  below.
 \begin{enumerate}
\item[In\ \ :]  $\boldsymbol{\lambda}\in\mathbb{F}^n$, which determines a Newton basis $\boldsymbol{N}_{\boldsymbol{\lambda}}(x)$ of $\mathbb{F}_{n}[x]$;\\ 
$F,G\in\mathbb{F}[x]$ in Newton basis $\boldsymbol{N}_{\boldsymbol{\lambda}}(x)$ with degrees $n$ and $m$ respectively, where $n\ge m$, i.e.,
\begin{eqnarray*}
F\left ( x \right ) =\sum_{i=0}^{n} a_{i} N_{i}(x),\quad G\left ( x \right ) =\sum_{i=0}^{m} b_{i} N_{i }(x) .
\end{eqnarray*}
\item[Out:] the B\'ezout matrix $\boldsymbol{B}_{\tilde{\boldsymbol{N}}_{\boldsymbol{\lambda}}}$ of $F$ and $G$ in $\tilde{\boldsymbol{N}}_{\boldsymbol{\lambda}}$.
\end{enumerate}

\begin{enumerate}[Step $1$.]
\item Convert $F$ and $G$ into their equivalent expressions $F'$ and $G'$ in monomial basis $\boldsymbol{P}$;
\item Call an algorithm from \cite[Section 3.3]{2002_Chionh_Zhang_Goldman}
for computing the B\'ezout matrix of $F'$ and $G'$ in  $\tilde{\boldsymbol{P}}$, resulting a matrix $\boldsymbol{B}_{\tilde{\boldsymbol{P}}}$;
\item Convert the matrix $\boldsymbol{B}_{\tilde{\boldsymbol{P}}}$ into the B\'ezout matrix $\boldsymbol{B}_{\tilde{\boldsymbol{N}}_{\boldsymbol{\lambda}}}$ in $\tilde{\boldsymbol{N}}_{\boldsymbol{\lambda}}$.
\end{enumerate}
It should be pointed out that  the algorithm from \cite[Section 3.3]{2002_Chionh_Zhang_Goldman}
is, to the best of our knowledge, the most efficient one for computing B\'ezout matrices in monomial basis.

\subsection{Complexity}\label{sub:complexity}

By Proposition \ref{prop:algorithm}-(3), the algorithm ${\sf BezNewton\_preserving}$ requires $\frac{n(3n+1)}{2}$  (i.e., $\mathcal{O} \left ( n^{2} \right )$) multiplications and $2n^2-n$ (i.e., $\mathcal{O} \left ( n^{2} \right )$) additions.

Now we analyze the complexity of the algorithm  ${\sf BezNewton\_trans}$ in each step. 
\begin{enumerate}[(1)]
\item In Step 1, the conversion of the input Newton polynomials into polynomials in monomial basis requires two multiplications of an $(n+1)\times(n+1)$ (or $(m+1)\times(m+1))$ upper-triangular transition matrix and a coefficient vector of length $n+1$ (or $m+1$), whose computational complexity is  $\left((n+1)(n+2)+(m+1)(m+2)\right)/2$ multiplications and $\left(n(n+1)+m(m+1)\right)/{2}$ additions. 

\item In Step 2, the computational complexity for computing the  B\'ezout matrix of the two standard polynomials with degrees $n$ and $m$ respectively
is  $n^{2}+n$ multiplications and $n^{2}$ additions \cite[Table 1]{2002_Chionh_Zhang_Goldman}.

\item The conversion of the B\'ezout matrix in monomial basis to that in Newton basis involves two multiplications of an $n\times n$ triangular matrix and an $n\times n$ full matrix, which  requires $n^{2}(n+1)$ multiplications and $n^{2}(n-1)$ additions.
\end{enumerate}

In summary, the total cost of the algorithm  ${\sf BezNewton\_trans}$ is
\begin{itemize} 
\item \# of multiplications:$\dfrac{(2n^{3}+5n^{2}+5n)+(m^2+3m+4)}{2}$ (i.e., $\mathcal{O} \left ( n^{3} \right )$ since $m\le n$), and   
\item \# of additions: $\dfrac{(2n^{3}+n^{2}+n)+(m^2+m)}{2}$ (i.e., $\mathcal{O} \left ( n^{3} \right )$ since $m\le n$).
\end{itemize}

The discussion above can be summarized into Table \ref{tab:complexity} below.

\begin{table}[htbp]
        \begin{center}
                \caption{The computational complexity of ${\sf BezNewton\_preserving}$  and   ${\sf BezNewton\_trans}$ }\label{tab:complexity}
                \begin{tabular}{ccc}
                        \hline    & ${\sf BezNewton\_preserving}$ &  ${\sf BezNewton\_trans}$ \\
                        \hline  $\#$  of mul. & $\mathcal{O}(n^{2})$ &  $\mathcal{O}(n^{3})$ \\
                        \hline  $\#$  of add. &$\mathcal{O}(n^{2})$& $\mathcal{O}(n^{3})$ \\
                        \hline
                \end{tabular}
        \end{center}
\end{table}   
                                                  
\subsection{Performance}\label{ssec:performance}
To compare the practical performance of the algorithms ${\sf BezNewton\_preserving}$ and algorithm ${\sf BezNewton\_trans}$,  we carry out a series of experiments.
These experiments are performed in Maple 2022 on a PC equipped with a CPU of Intel(R) Core(TM) i5-9300H (2.40GHz) and a RAM of 16.0GB.
The examples used in the experiment are classified into two groups: one with constant coefficients and the other with parametric coefficients. For polynomials in the first group, the coefficients are randomly generated and so are $\boldsymbol{\lambda}$'s. For polynomials in the second group, we set $\boldsymbol{\lambda}$'s to be random numeric vectors and the coefficients are formal ones (i.e., $a_i$'s and $b_j$'s).
Both the program and the test examples can be accessed via the link https://github.com/JYangMATH/NewtonBez.git. The  computational results are shown in Table \ref{tab:performance1} and Table \ref{tab:performance2}, respectively, where $t_{\sf preserving}$ and $t_{\sf trans}$ are the time cost charged by the algorithms ${\sf BezNewton\_preserving}$ and $\sf{BezNewton\_trans}$, respectively.

\begin{table}[htbp]
\begin{center}
\caption{Comparison on the time cost (in seconds) of ${\sf BezNewton\_preserving}$ and $\sf{BezNewton\_trans}$ for polynomials with constant coefficients.}
\label{tab:performance1}
\begin{tabular}{ccrrr}
\hline Degree&Matrix size&$t_{\sf preserving}$&$t_{\sf trans}$& $\frac{t_{\sf trans}}{t_{\sf preserving}}$\\
\hline
$(102,98)$& $102\times102$ & $0.015$ &$2.390$ & 159.33\\
$(135,127)$& $135\times135$ & $0.031$ &$6.078$ & 196.06\\
$(168,163)$& $168\times168$ & $0.046$ &$11.796$ & 256.43\\
$(201,199)$& $201\times201$ & $0.078$ &$24.015$ & 307.88\\
$(234,228)$& $234\times234$ & $0.109$ &$130.875$  & 1200.69\\
$(267,265)$& $267\times267$ & $0.156$ &$333.500$  & 2137.82\\
$(300,290)$& $300\times300$ & $0.171$ &$557.125$  & 3258.04\\
$(333,326)$& $333\times333$ & $0.203$ &$989.062$  & 4872.23\\
$(366,359)$& $366\times366$ & $0.265$ &$1464.156$  & 5525.12\\
$(399,391)$& $399\times399$ & $0.296$ &$2283.281$  & 7713.79\\
$(432,432)$& $432\times432$ & $0.406$ &$3511.890$  & 8649.98\\
$(500,500)$& $500\times500$ & $0.515$ &$5355.609$  & 10748.76\\
\hline
\end{tabular}
\end{center}
\end{table}

 \begin{table}
 \begin{center}
 \caption{Comparison on the time cost (in seconds) of ${\sf BezNewton\_preserving}$ and $\sf{BezNewton\_trans}$ for polynomials  with parametric coefficients}
 \label{tab:performance2}
 \begin{tabular}{ccrrr}
 \hline Degree&Matrix size&$t_{\sf preserving}$&$t_{\sf trans}$& $\frac{t_{\sf trans}}{t_{\sf preserving}}$\\
 \hline
 $(35,35)$& $35\times35$ & $0.031$ &$1.656$ & 53.42\\
 $(40,40)$& $40\times40$ & $0.062$ &$2.687$ & 43.34\\
 $(45,45)$& $45\times45$ & $0.078$ &$6.390$ & 81.92\\
 $(50,50)$& $50\times50$ & $0.140$ &$16.515$ & 117.96\\
 $(55,55)$& $55\times55$ & $0.156$ &$21.046$ & 134.91\\
 $(60,60)$& $60\times60$ & $0.296$ &$38.531$ & 130.17\\
 $(65,65)$& $65\times65$ & $0.421$ &$60.000$ & 142.52\\
 $(70,70)$& $70\times70$ & $0.609$ &$90.343$ & 148.37\\
 $(75,75)$& $75\times75$ & $0.718$ &$140.796$ & 196.09\\
 $(80,80)$& $80\times80$ & $0.968$ &$264.593$  & 273.34\\
 $(85,85)$& $85\times85$ & $1.218$ &$326.984$  & 268.46\\
 $(90,90)$& $90\times90$ & $1.578$ &$614.593$  & 389.48\\
 \hline
 \end{tabular}
 \end{center}
 \end{table}

From Tables \ref{tab:performance1} and \ref{tab:performance2}, we make the following observations.
\begin{enumerate}[(1)]
\item For both cases, we have $t_{\sf recursive}<<t_{\sf trans}$, indicating that the algorithm ${\sf BezNewton\_preserving}$ behaves significantly better than the algorithm $\sf{BezNewton\_trans}$. 
Even for numeric polynomials of degree $500$, the recursive approach only costs less than one second while the basis-transformation-based approach costs almost three hours.

\item 
The ratio $t_{\sf trans}/t_{\sf preserving}$ increases
as the degree of the input polynomial increases, which shows that the algorithm ${\sf BezNewton\_preserving}$ has better scalability than $\sf{BezNewton\_trans}$. This agrees with the magnitude of their computational complexity.
\end{enumerate}

We attribute the dominant performance of the proposed algorithm to the avoidance of basis transformation, which are computationally intensive. Another benefit brought by the avoidance of basis transformation is that an inflation on the size of coefficients, which  makes the computation unstable, will not occur.

\section{Application}\label{sec:application}

In this section, we show an application of the proposed algorithm in computing the confederate resultant matrix of Newton polynomials, which can be used to formulate subresultant polynomials in Newton basis \cite{2022_Wang_Yang}.
\subsection{Problem formulation}
B\'ezout matrix in  monomial basis can be used to construct companion resultant matrices in monomial basis, by making use of the relationship between them \cite{1972_Barnett}. In \cite{2001_Yang}, Yang revealed that the relationship between the two matrices formulated in monomial basis can be extended to that in general basis with a specialization to be Newton basis. Thus we can use the B\'ezout matrix in Newton polynomials to construct the confederate resultant matrix of Newton polynomials.

We begin by a review on the concept of the well known companion matrix.
\begin{definition}[\cite{1933_MacDuffee}]\label{def:CP}
Let $F=a_{n}x^{n}+\cdots+a_{1}x+a_{0}\in \mathbb{F} \left [ x \right ] $. The companion matrix of $F$ with respect to $x$ is defined as an $n\times n$ matrix $\boldsymbol{C}(F)$ such that
\begin{equation}\label{eq:CP}
        x\cdot \tilde{\boldsymbol{P}}(x)\equiv _{F} \boldsymbol{C}(F)\tilde{\boldsymbol{P}}(x)
\end{equation}
where  $\tilde{\boldsymbol{P}}(x)=(x^{n-1},\ldots,x^{0})$. More explicitly,
\begin{align*}
\boldsymbol{C}(F)=\begin{pmatrix}
        -\dfrac{a_{n-1}}{a_{n}}  & -\dfrac{a_{n-2}}{a_{n}} & \cdots & -\dfrac{a_{1}}{a_{n}} & -\dfrac{a_{0}}{a_{n}}\\
        1 & 0 &  &  &   \\
         & 1 & \ddots &   &   \\
         & & \ddots  & 0&   \\
         &  &    & 1 & 0  
\end{pmatrix}
\end{align*}
The matrix $G(\boldsymbol{C}(F))$ is called the companion resultant matrix of $F$ and $G$ with respect to $x$. 
\end{definition}

The companion resultant matrix can be used to formulate generalized subresultant polynomials of several univariate polynomials, called Barnett-type subresultant polynomials (see \cite{2021_Hong_Yang}).
The following Barnett formula provides a practical way to compute the
companion resultant matrix.
\begin{proposition}[Barnett formula \cite{1983_Barnett}]
Given  $F$ and $G\in\mathbb{F}[x]$, we have
$$G(\boldsymbol{C}(F))=\boldsymbol{B}_{\tilde{\boldsymbol{P}}}(F,1)^{-1}\boldsymbol{B}_{\tilde{\boldsymbol{P}}}(F,G)$$
\end{proposition}

In \cite{1979_Maroulas_Barnett}, Maroulas and Barnett generalized the concept of companion matrix to confederate matrix in general basis. The main idea is to maintain the relationship defined by \eqref{eq:CP}, which captures the most essential property of companion matrices. Later Yang showed that the Barnett formula can be extended to the case of confederate matrix and B\'ezout matrix in general basis.
Since Newton basis is a specialization of general basis and is what we are concerned with in the current paper, we recall the concept of confederate matrix in Newton basis below. 

\begin{definition}[\cite{2006_Amiraslani_Corless,1979_Maroulas_Barnett}]\label{def:CN}
Let $\boldsymbol{\lambda}\in\mathbb{F}^{n}$ and $\boldsymbol{N}_{\boldsymbol{\lambda}}(x)$ be the Newton basis associated with $\boldsymbol{\lambda}$. Let $F=a_{n}N_{n}(x)+\cdots+a_{1}N_{1}(x)+a_{0}N_{0}(x)\in \mathbb{F}\left [ x \right ] $ and $\tilde{\boldsymbol{N}}_{\boldsymbol{\lambda}}$ be obtained by truncating the first polynomial in $\boldsymbol{N}_{\boldsymbol{\lambda}}$. Then the confederate matrix of $F$ with respect to $x$ in $\tilde{\boldsymbol{N}}_{\boldsymbol{\lambda}}(x)$ is defined as an $n\times n$ matrix $\boldsymbol{C}_{\tilde{\boldsymbol{N}}_{\boldsymbol{\lambda}}}(F)$ such that
\[
 x\cdot \tilde{\boldsymbol{N}}_{\boldsymbol{\lambda}}(x)\equiv _{F} \boldsymbol{C}_{\tilde{\boldsymbol{N}}_{\boldsymbol{\lambda}}}(F)\tilde{\boldsymbol{N}}_{\boldsymbol{\lambda}}(x)
\]
More explicitly,
\begin{align*}
\boldsymbol{C}_{\tilde{\boldsymbol{N}}_{\boldsymbol{\lambda}}}(F)=\,\begin{pmatrix}
        \dfrac{a_{n}\lambda_{n}-a_{n-1} }{a_{n}}  & \dfrac{-a_{n-2} }{a_{n}}  &\cdots   &  \dfrac{-a_{0} }{a_{n}}  \\
        1 & \lambda _{n-1} &  &   \\
        & \ddots & \ddots  &   \\
        &  & 1 & \lambda _{1}
\end{pmatrix}
\end{align*}
The matrix $G(\boldsymbol{C}_{\tilde{\boldsymbol{N}}_{\boldsymbol{\lambda}}}(F))$ is called the confederate resultant matrix of $F$ and $G$ with respect to $x$ in  $\tilde{\boldsymbol{N}}_{\boldsymbol{\lambda}}$.
\end{definition}

One may easily derive the relationship   between 
 $G(\boldsymbol{C}(F))$ and  $G(\boldsymbol{C}_{\tilde{\boldsymbol{N}}_{\boldsymbol{\lambda}}}(F)).$ \begin{proposition}\label{prop:relation_CP_CN}
With the above settings, we have
$G\left(\boldsymbol{C}_{\tilde{\boldsymbol{N}}_{\boldsymbol{\lambda}}}(F)\right)=U^{-1}\cdot G\left(\boldsymbol{C}(F)\right)\cdot U$ where $U$ is the transition matric from $\tilde{\boldsymbol{P}}$ to $\tilde{\boldsymbol{N}}_{\boldsymbol{\lambda}}$.
\end{proposition}

Zhang extended the Barnett formula to the case of confederate matrices and B\'ezout matrix in \cite{2001_Yang}, which is stated below.
\begin{proposition}[\cite{2001_Yang}]\label{prop:gen_Barnett_formula}
With the above settings, we have
$$G\left(\boldsymbol{C}_{\tilde{\boldsymbol{N}}_{\boldsymbol{\lambda}}}(F)\right) =\boldsymbol{B}_{\tilde{\boldsymbol{N}}_{\boldsymbol{\lambda}}}(F,1)^{-1}\boldsymbol{B}_{\tilde{\boldsymbol{N}}_{\boldsymbol{\lambda}}}(F,G)$$
\end{proposition}

Similar to the companion resultant matrix, the confederate resultant matrix can also be used to formulate generalized subresultant polynomials of several univariate polynomials in Newton polynomials (see \cite{2022_Wang_Yang}).
Hence we need a practical way to compute the
confederate resultant matrix in Newton basis.

Now we are ready to make a formal presentation of the problem to be studied in this section.
\begin{problem}
The problem of computing confederate resultant matrix  of two polynomials in Newton basis can be stated as follows.
\begin{enumerate}
\item [ In\ \ :]  $\boldsymbol{\lambda}\in\mathbb{F}^n$, which determines a Newton basis $\boldsymbol{N}_{\boldsymbol{\lambda}}(x)$ of $\mathbb{F}_{n}[x]$;\\ 
$F,G\in\mathbb{F}[x]$ in Newton basis $\boldsymbol{N}_{\boldsymbol{\lambda}}(x)$ with degrees $n$ and $m$ respectively, where $n\ge m$, i.e.,
\begin{eqnarray*}
F\left ( x \right ) =\sum_{i=0}^{n} a_{i} N_{i}(x),\quad G\left ( x \right ) =\sum_{i=0}^{m} b_{i} N_{i }(x) .
\end{eqnarray*}
\item[Out:] the confederate  resultant matrix $G(\boldsymbol{C}_{\tilde{\boldsymbol{N}}_{\boldsymbol{\lambda}}}(F))$  of $F,G$  in the Newton basis $\tilde{\boldsymbol{N}}_{\boldsymbol{\lambda}}$.
\end{enumerate}
\end{problem} 

\subsection{Three approaches for computing confederate resultant matrices}\label{ssec:3approaches}
In this subsection, we propose three methods for computing confederate resultant matrices in Newton basis. 
\begin{enumerate}[(1)]
\item Approach A: matrix polynomial evaluation + Horner algorithm

Approach A computes confederate resultant matrices in a straightforward way. More explicitly, we evaluate the value of $G(x)$ when $x=\boldsymbol{C}_{\tilde{\boldsymbol{N}}_{\boldsymbol{\lambda}}}(F)$, which is a matrix polynomial evaluation. Hence we can use the well known Horner algorithm for polynomial evaluating to improve the computational efficiency.    

\item Approach B: basis transformation + Barnett formula

\begin{enumerate}[Step {B}$1$.]
\item Convert $F$ and $G$ into their equivalent expressions $F'$ and $G'$ in monomial basis $\boldsymbol{P}$;
\item Call an algorithm from \cite[Section 3.3]{2002_Chionh_Zhang_Goldman}
to compute the B\'ezout matrix of $F'$ and $G'$, and that of $F'$ and $1$, in  $\tilde{\boldsymbol{P}}$, resulting two matrices $\boldsymbol{B}_{\tilde{\boldsymbol{P}}}(F',G')$ and $\boldsymbol{B}_{\tilde{\boldsymbol{P}}}(F',1)$;

\item Compute $G(\boldsymbol{C}(F))$ by multiplying the inverse of $\boldsymbol{B}_{\tilde{\boldsymbol{P}}}(F',1)$ and $\boldsymbol{B}_{\tilde{\boldsymbol{P}}}(F',G')$.

\item Convert the companion resultant matrix $G(\boldsymbol{C}(F))$ to the confederate resultant matrix $G(\boldsymbol{C}_{\tilde{\boldsymbol{N}}_{\boldsymbol{\lambda}}}(F))$  of $F,G$  in $\tilde{\boldsymbol{N}}_{\boldsymbol{\lambda}}$ by Proposition \ref{prop:relation_CP_CN}.
\end{enumerate}

\item Approach C: generalized Barnett formula

\begin{enumerate}[Step {C}$1$.]
\item Call the algorithm ${\sf BezNewton\_preserving}$ for computing the B\'ezout matrices of $F$ and $G$, and that of $F$ and $1$, in the Newton basis $\tilde{\boldsymbol{N}}_{\boldsymbol{\lambda}}$, resulting two matrices $\boldsymbol{B}_{\tilde{\boldsymbol{N}}_{\boldsymbol{\lambda}}}(F,G)$ and $\boldsymbol{B}_{\tilde{\boldsymbol{N}}_{\boldsymbol{\lambda}}}(F,1)$;

\item Compute $G\left(\boldsymbol{C}_{\tilde{\boldsymbol{N}}_{\boldsymbol{\lambda}}}(F)\right) $ by multiplying the inverse of $\boldsymbol{B}_{\tilde{\boldsymbol{N}}_{\boldsymbol{\lambda}}}(F,1)$ and $\boldsymbol{B}_{\tilde{\boldsymbol{N}}_{\boldsymbol{\lambda}}}(F,G)$.
\end{enumerate}
\end{enumerate}
In the following subsection, we will show that Approach C which employs the algorithm ${\sf BezNewton\_preserving}$ proposed in the current paper has the best performance.

\subsection{Comparison}
In this section, we compare the three approaches presented in Subsection \ref{ssec:3approaches}. The comparison is carried out from two aspects, i.e., computational complexity and practical performance.

\subsubsection{Complexity}

We first give a detailed complexity analysis for each of the three approaches presented above. 

In Approach A, the main cost is the evaluation of a matrix polynomial. The computation of $G\left(\boldsymbol{C}_{\tilde{\boldsymbol{N}}_{\boldsymbol{\lambda}}}(F)\right) $ with the Horner algorithm requires $m$ matrix multiplications and $m$ matrix additions where the involved matrices are of size $n\times n$. Since each matrix multiplication needs $\mathcal{O}(n^{3})$ multiplications and  $\mathcal{O}(n^{3})$ additions. Hence the total cost of Approach A is $\mathcal{O}(mn^{3})$ multiplications and  $\mathcal{O}(mn^{3})$ additions.

Approach B has four steps and we analyze the complexity for each step.
\begin{enumerate}[(1)]
\item In Step B1, 
the conversion of the input Newton polynomials into polynomials in monomial basis requires two multiplications of an $(n+1)\times(n+1)$ (or $(m+1)\times(m+1))$ upper-triangular transition matrix and a coefficient vector of length $n+1$ (or $m+1$). Since $m\le n$, the total cost in  Step B1 is   $\mathcal{O} \left ( n^{2} \right)$ multiplications and $\mathcal{O} \left ( n^{2} \right)$ additions. 

\item Step B2 only involves the computation of two B\'ezout matrices in monomial basis. By \cite[Table 1]{2002_Chionh_Zhang_Goldman}, each of them charges    $\mathcal{O} \left ( n^{2} \right)$ multiplications and $\mathcal{O} \left ( n^{2} \right)$ additions.

\item In Step B3, we need to compute the inverse of an $n\times n$ matrix and the multiplication of two $n\times n$ matrices and each of them requires     $\mathcal{O} \left ( n^{3} \right)$ multiplications and $\mathcal{O} \left ( n^{3} \right)$ additions.

\item Step B4 requires two multiplications of $n\times n$ matrices, which needs      $\mathcal{O} \left ( n^{3} \right)$ multiplications and $\mathcal{O} \left ( n^{3} \right)$ additions.
\end{enumerate}
Altogether, the cost of Approach B is      $\mathcal{O} \left ( n^{3} \right)$ multiplications and $\mathcal{O} \left ( n^{3} \right)$ additions.

In Approach C, Step C1 involves the computation of two B\'ezout matrices in Newton basis. From Table \ref{tab:complexity}, each of them charges    $\mathcal{O} \left ( n^{2} \right)$ multiplications and $\mathcal{O} \left ( n^{2} \right)$ additions. Step C2 requires one matrix inversion and one matrix multiplication where the matrices involved are all of order $n\times n$. 
Each of them require $\mathcal{O}(n^{3})$ multiplications and $\mathcal{O}(n^{3})$ additions. Thus, the total cost of Approach C is $ \mathcal{O} \left ( n^{3} \right ) $  multiplications and $ \mathcal{O} \left ( n^{3} \right )$ additions.

We summarize the above discussion in Table \ref{tab:complexity_2}.
\begin{table}[htbp]
        \begin{center}
                \caption{The computational complexity of Approaches A, B and C}\label{tab:complexity_2}
                \begin{tabular}{cccc}
                        \hline    &Approach A  &Approach  B &Approach C\\
                        \hline  $\#$  of mul. & $\mathcal{O}(mn^3)$ &  $\mathcal{O}(n^3)$ & $\mathcal{O}(n^3)$\\
                        \hline  $\#$  of add. &$\mathcal{O}(mn^3)$& $\mathcal{O}(n^3)$ & $\mathcal{O}(n^3)$\\
                        \hline
                \end{tabular}
        \end{center}
\end{table}
\subsubsection{Performance}
In order to evaluate the practical performance of the three approaches, we conduct a series of experiments on numeric and parametric polynomials. The specification for the input polynomials is the same as experiments in Subsection \ref{ssec:performance}.
These experiments are performed in Maple 2022 on a PC equipped with a CPU of Intel(R) Core(TM) i5-9300H(2.40GHz) and a RAM of 16.0GB.
Both the program and the test examples can be accessed via the link https://github.com/JYangMATH/NewtonBez.git. The experimental results are reported in Tables \ref{tab:performance3} (for numeric polynomials) and \ref{tab:performance4} (for parametric polynomials) where  $t_\text{A}$, $t_\text{B}$, $t_\text{C}$ are the time costs of Approaches A, B, and C, respectively.

\begin{table}[htbp]
\begin{center}
\caption{The time cost (in sec.) of Approaches A, B, and C for numeric polynomials }
\label{tab:performance3}
\begin{tabular}{ccccccc}
\hline Degree&Matrix size&$t_\text{A}$&$t_\text{B}$&$t_\text{C}$& $\frac{t_\text{A}}{t_\text{C}}$&$\frac{t_\text{B}}{t_\text{C}}$\\
\hline
$(50,50)$& $50\times50$ & $0.750$ &$0.468$ & $0.156$ &4.81& 3.00\\
$(70,70)$& $70\times70$ & $2.187$ &$1.312$ & $0.375$&5.83& 3.50\\
$(90,90)$& $90\times90$ & $8.234$ &$3.203$ & $0.921$&8.94&3.48\\
$(110,110)$& $110\times110$ & $22.593$ &$7.031$ & 2.390 & 9.45 & 2.94\\
$(130,130)$& $130\times130 $ & $52.890$ &$11.078$ &3.250&16.27& 3.41\\
$(150,150)$& $150\times150$ & $107.203$ &$20.687$ & 6.593&16.26& 3.14\\
$(170,170)$& $170\times170$ & $166.587$ &$19.796$ & 5.125&32.50&3.86\\
$(190,190)$& $190\times190$ & $303.796$ &$56.218$ & 15.843&19.18&3.55\\
$(210,210)$& $210\times210$ & $680.140$ &$83.500$ & 23.078&29.47& 3.62\\
$(230,230)$& $230\times230$ & $720.156$ &$186.406$ & 31.843&22.62& 5.85\\
$(250,250)$& $250\times250$ & $993.953$ &$335.500$  & 49.718&19.99&6.75\\
$(270,270)$& $270\times270$ & $5721.625$ &$436.125$  & 51.765&110.53&8.43 \\
\hline
\end{tabular}
\end{center}
\end{table}

\begin{table}[htbp]
        \begin{center}
\caption{The time cost (in sec.) of Approaches A, B, and C for parametric polynomials}
\label{tab:performance4}
\begin{tabular}{ccccccc}
\hline Degree&Matrix size&$t_{\text{A}}$&$t_{\text{B}}$&$t_{\text{C}}$& $\frac{t_{\text{A}}}{t_{\text{C}}}$&$\frac{t_{\text{B}}}{t_{\text{C}}}$\\
\hline
$(16,16)$& $16\times16$ & $7.000$ &$0.046$ & $0.031$&$225.80$&$1.48$\\
$(17,17)$& $17\times17$ & $16.609$ &$0.093$ & $0.062$&$267.89$&$1.50$\\
$(18,18)$& $18\times18$ & $40.812$ &$0.109$ & $0.093$&$438.84$&$1.17$\\
$(19,19)$& $19\times19$ & $112.031$ &$0.187$ & $0.140$&$800.22$&$1.36$\\
$(20,20)$& $20\times20$ & $251.734$ &$0.281$ & $0.187$&$1346.17$&$1.50$\\
$(21,21)$& $21\times21$ & $587.687$ &$0.250$ & $0.218$&$3142.71$&$1.15$\\
$(22,22)$& $22\times22$ & $1487.531$ &$0.437$ & $0.390$&$3739.82$&$1.12$\\
$(25,25)$& $23\times23$ & $>6000$ &$1.468$ & $1.062$&$>5649.72$&$1.38$\\
$(30,30)$& $24\times24$ & $>6000$ &$6.250$ & $3.375$&$>5649.72$&$1.85$\\
$(35,35)$& $25\times25$ & $>6000$ &$34.390$ & $17.640$&$>5649.72$&$1.95$\\
$(40,40)$& $26\times26$ & $>6000$ &$155.921$ & $83.453$&$>5649.72$&$1.89$\\
$(45,45)$& $27\times27$ & $>6000$ &$692.234$ & $349.500$&$>5649.72$&$1.98$\\
\hline
                \end{tabular}
        \end{center}
\end{table}

From Tables \ref{tab:performance3} and \ref{tab:performance4} , we make the following observations:
\begin{enumerate}[(1)]
\item $t_\text{C}<t_\text{B}<<t_\text{A}$, indicating that Approach C behaves better than the other two and Approach A is the worst. 

\item The ratio $t_\text{B}/t_\text{C}$ exhibits a slow increase as the degree increases, which means  Approach C has a better scalability than Approach B though their complexity is at the same level. We suspect the underlying reason is the magnitude of the entries in the matrices representing the involved basis transformations are increasing as the degree of the input polynomials becomes higher.       

\item 
The ratio $t_\text{A}/t_\text{C}$ increases
as the degree of the input polynomial increases, which shows that Approach C has a much better scalability than Approach A. This is expected from the computational analysis.
\end{enumerate}

\section{Conclusion and Perspectives}
\label{sec:conclusion}

In this paper, we consider the problem of constructing
the B\'ezout matrix for Newton polynomials. It is required that the output matrix is also formulated in the given basis used to define the input polynomials. After investigating the structure hidden in B\'ezout matrices in Newton basis, we design an efficient algorithm for constructing such a matrix. It should be pointed out that the proposed algorithm does not require basis transformation and thus can avoid the heavy computational cost and numerical instability caused by transformation. We also show an application of the proposed algorithm in the construction of  confederate resultant matrix for Newton polynomials. Experimental results show that the proposed methods display a significantly better behavior compared with the basis-transformation-based one.

Another application of B\'ezout matrix in Newton basis is to formulate subresultant polynomials (e.g., \cite{2022_Wang_Yang,2023_Yang_Yang}) for Newton polynomials. One merit of the subresultant polynomials formulated with  B\'ezout matrix
is that the basis is preserved. However, there is still a lack of works on a deep understanding of  the structure of B\'ezout matrix from which  the computation of  subresultant polynomials in Newton basis can be accelerated. This could be an interesting topic worthy of further investigation in the future.

\medskip
\noindent{\bf Acknowledgements.} This research was supported by  the National Natural Science Foundation of China (Grant Nos.: 12326353 and 12261010), the Natural Science
Foundation of Guangxi (Grant No.: 2023GXNSFBA026019), and the Natural Science
Cultivation Project of GXMZU (Grant No.: 2022MDKJ001).


\end{document}